\documentclass[multphys,vecphys]{promult}

\usepackage{makeidx}     
\usepackage{graphicx}    
\usepackage{multicol}    

\makeindex             

\begin{document}

\title*{Gamma Ray Bursts in the Era of Rapid Followup}
\author{C.G.~Mundell\inst{1}, C.~Guidorzi\inst{2,1}\, and I.A.~Steele\inst{1}  on behalf of the Liverpool GRB team}
\authorrunning{Mundell, Guidorzi \& Steele} 
\institute{Astrophysics Research Institute, Liverpool John Moores University, Twelve Quays House, Birkenhead, CH41 1LD, U.K. cgm@astro.livjm.ac.uk
\and Physics Department, University of Ferrara, via Saragat 1, 44122, Ferrara, Italy. guidorzi@fe.infn.it} 
\maketitle

We present a status report on the study of gamma-ray bursts (GRB) in the era of rapid follow-up using the world's largest robotic optical telescopes - the 2-m Liverpool and Faulkes telescopes. Within the context of key unsolved issues in GRB physics, we describe (1) our innovative software that allows real-time automatic analysis and interpretation of GRB light curves, (2) the novel instrumentation that allows unique types of observations  (in particular, early time polarisation measurements), (3) the key science questions and discoveries to which robotic observations are ideally suited, concluding with a summary of current understanding of GRB physics provided by combining rapid optical observations with simultaneous observations at other wavelengths.


\section{Introduction}
\label{sec:intro}

Gamma-Ray Bursts (GRBs) are the most powerful explosions in the Universe and, arguably, represent the most significant new astrophysical phenomenon since the discovery of quasars and pulsars. As their name suggests, GRBs are detected as brief, intense and totally unpredictable flash of high-energy gamma rays, thought to be produced during the core collapse of massive stars (long-soft bursts, T$_\gamma$$>$2~s) or the merger of two compact objects such as two neutron stars or a neutron star and a stellar-mass black hole (short-hard bursts, T$_\gamma$$<$2~s). Although discovered through their $\gamma$-ray emission~\cite{klebesadel73}, they are now known to emit non-thermal radiation detectable across the electromagnetic spectrum \cite{costa97,vanparadijs97,frail97}. However, despite their enormous luminosity, their unpredictability and short duration limit rapid, accurate localisation and observability with traditional telescopes. Consequently, new ground and space-based facilities have been developed over the past decade; dedicated satellites optimised for GRB detection and followup, such as Swift~\cite{gehrels04}, are revolutionizing GRB studies by locating $\sim$100 bursts per year with $\gamma$-ray positions accurate to $\sim$3$'$ and X-ray positions accurate to 5$"$ within seconds or minutes of the burst. Here we describe the automatic ground-based follow-up of GRBs with the world's largest robotic optical telescopes that use intelligent software and innovative instruments. 

\paragraph{The Era of Rapid Follow-up: Predictions and Outcomes }

Before the launch of current satellites such as Swift, Integral and Fermi, significant progress in understanding GRBs had been made since their discovery, in particular the general $\gamma$ and X-ray properties.
The first crucial step in disseminating real-time GRB positions to ground observers was triggered
by BATSE on the CGRO~\cite{paciesas99} through the GRB Coordinates Network (GCN) \cite{barthelmy08}
via internet socket connection (no humans-in-the-loop).
This drove development of the first generation of wide-field robotic followup ground-based facilities, such as  GROCSE, ROTSE, and LOTIS, culminating with the discovery of the optical flash associated with
GRB~990123 \cite{akerlof99}. BATSE provided an invaluable catalogue of prompt $\gamma$-ray profiles, whose isotropic sky distribution and inhomogeneous intensity distribution suggested a cosmological origin \cite{paciesas99}, and BeppoSAX \cite{boella97} revolutionised the cosmological study of GRBs by providing sub-arcmin ($\sim$50$"$) localisation of X-ray afterglows that enabled late-time ($\sim$ hours) optical followup with traditional ground-based telescopes and redshift determinations. Collimation of the ejecta (i.e. jets)  was inferred from temporal breaks - steepening - of optical light curves at $\sim$1 day post-burst and the concept of a universal central engine and the use of GRBs as standardisable cosmological candles was introduced \cite{frail01,ghirlanda04}. 

\begin{figure}
\centering
\includegraphics[height=6cm]{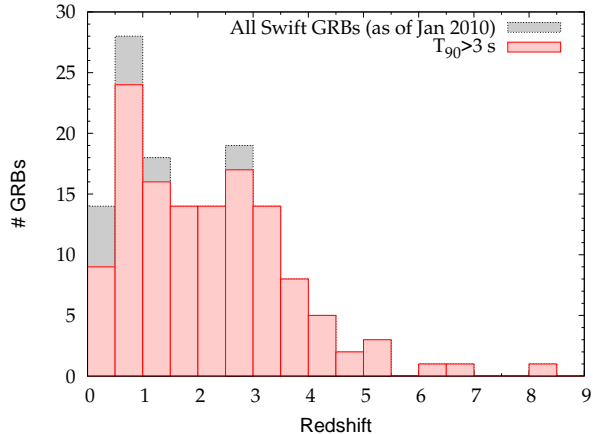}
\caption{Redshift distribution of {\em Swift} GRBs detected to-date.}
\label{fig:z}       
\end{figure}

The possibility for great advances with the launch of Swift was fully recognised. Optical counterparts were expected to be found for all GRBs with many GRBs expected to exhibit bright optical flashes from reverse shock emission at early times, similar to GRB 990123 \cite{akerlof99}. An increase in the number of GRBs detected would lead to many jet breaks being identified, short GRBs would be easily observed and understood and identification of GRBs at very high redshift would be routine. 
Instead, 50\% of GRBs remain optically dark, despite deep, rapid followup \cite{melandri08,cenko09,oates09,yost07}; there is a dearth of bright reverse-shock optical emission \cite{roming06}; light curves are complex in all bands with a variety of chromatic and achromatic breaks and flares observed (e.g., \cite{tagliaferri05,burrows05,obrien06,chincarini07,falcone07,melandri08,klotz09,rykoff09}). Jet breaks have proven elusive, short bursts remain technically challenging \cite{graham09} and only 3 GRBs have been identified to lie a $z>6$  (Fig. \ref{fig:z}) \cite{kawai06,greiner09a,salvaterra09,tanvir09}.

\section{Robotic Follow-up and Intelligent Autonomy}
\label{sec:robots}

\begin{figure}
\centering
\includegraphics[height=6cm]{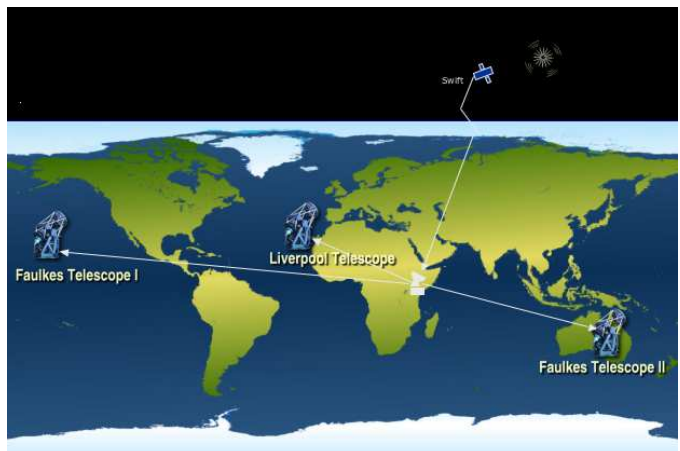}
%
%
\caption{The locations of the 2-m robotic Liverpool Telescope and its clones the Faulkes telescopes. GRB triggers from satellites drive prompt automatic follow up.}
\label{fig:map}       
\end{figure}

The field of GRB research is the most rapidly evolving topic in modern day astrophysics - driven primarily by technological innovation in both hardware and software; most notably is the need for rapid, intelligent and fully autonomous followup. Within the context of robotic telescopes devoted to searching for  optical counterparts to GRBs (e.g., [29--36]),\nocite{zerbi01,vestrand04,reichart05,yost06,klotz08,cenko06,bloom06} the Liverpool Telescope (LT) offers a unique combination of sensitivity, speed, instrument choice, and real-time reduction pipeline complexity and flexibility. The LT, owned and operated by Liverpool John Moores University (LJMU), has a 2-m diameter mirror, altitude-azimuth design, final focal ratio f/10, a comprehensive suite of instruments and a fully robotic control system. As shown in Figure \ref{fig:map}, it is sited at the Observatorio del Roque de los Muchachos in La Palma. Optimised for robotic follow-up of transient sources, the LT was designed to have a fully open enclosure, robustness to wind gusts and  a fast slew rate of 2$^\circ$/s; currently the fastest telescope for its size, it observes GRBs within 1$-$3~mins of receipt of a satellite alert. Although smaller  robotic telescopes slew more quickly, the LT is more sensitive to fainter bursts at early time. Most importantly,  the LT can perform unique early time polarisation measurements (see Section \ref{sec:magnets}).

\begin{table}
\centering
\caption{Current instrumentation on the Liverpool and Faulkes telescopes and associated GRB science goals; upcoming instrumentation is shown in italics.}
\label{tab:instruments}       
\begin{tabular}{lcl}
\hline\noalign{\smallskip}
{\bf Instrumentation} &~~& {\bf $\bullet$~Science Goals}\\
\noalign{\smallskip}\hline\noalign{\smallskip}
Optical Camera (FoV $\sim$5$'$) &~~& $\bullet$~Early multicolour light curves\\
(BVRi'z')  &~~& $\bullet$~Shock physics/ISM studies\\
(LT/FTN/FTS)  &~~& $\bullet$~Later-time light curves/jet breaks \\
                          &~~&$\bullet$~GRB-supernova connection\\
\hline 
RINGO2 polarimeter (FoV $\sim$4$'$)&~~&$\bullet$~Early-time polarisation studies\\
(LT only) &~~&~~~1\% polarisation at r$'$$<$ 17 mag in 1 min\\
               &~~&$\bullet$~Fundamental tests of magnetization\\
\hline
SupIRCam Infrared Camera (FoV $\sim$1$'$)&~~&$\bullet$~High-z 'naked' bursts\\
(LT only)&~~&$\bullet$~Low-z 'obscured' bursts\\
\hline
FRODOSpec IFU (FoV $\sim$11$"$)&~~&$\bullet$~Early evolution of circumburst medium\\
(LT only)\\
\hline
STILT (FoV 1$^o$/20$^o$/180$^o$)&~~&$\bullet$~Bright bursts/neutrino counterparts\\
(LT only)\\
\hline

{\em IO -- wide-field optical/IR imager}&~~&$\bullet$~Deep, simultaneous optical/IR light curves \\
~~~{\em (LT - 2010)}&~~& ~~~over 2$^o$ FoV\\
\noalign{\smallskip}\hline
\end{tabular}
\end{table}

\begin{figure}
\centering
\includegraphics[height=6cm]{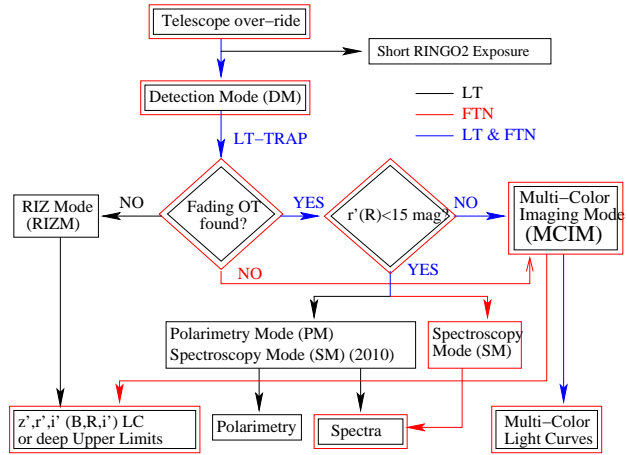}
\caption{Flow chart of the robotic GRB pipeline currently running on the Liverpool and Faulkes Telescopes (adapted from
\cite{guidorzi06a}).}
\label{fig:LT-TRAP}       
\end{figure}

The LT has five instrument ports: four folded and one straight-through, selected automatically within 30~s by a deployable rotating mirror in the Acquisition and Guidance (AG) box. Table \ref{tab:instruments} describes the instruments available and the related GRB science goals. All GRB follow-ups with the LT begin with RINGO polarimetry exposures before continuing with a sequence of 10-s R-band exposures that are used to automatically identify an optical counterpart, determine its characteristics and conduct subsequent optimized followup observations with the most appropriate instrument.  Guidorzi et al. \cite{guidorzi06a} describe the intelligent software logic - LT-TRAP - developed to perform the real-time analysis and follow-up.  Automatic multicolour light curves of optical transients brighter than R$\sim$19 mag are produced in real-time; transients brighter than R$\sim$15 mag trigger additional RINGO polarisation observations before continuing with multicolour imaging. Faint OTs or non-detections trigger deep exposures in red filters. The 2-m Faulkes telescopes\footnote{Now owned by Las Cumbres Observatory, are operated with support from the Dill Faulkes Educational Trust.}, clones of the LT,  provide additional sky coverage, operate the same intelligent LT-TRAP software, apart from small differences due to the different instruments mounted on each facility, and concentrate on multicolour optical imaging (Fig. \ref{fig:map} and Table \ref{tab:instruments}). Figure~\ref{fig:LT-TRAP} displays the flow chart of the LT-TRAP.

\section{Characteristics of Multiwavelength Light Curves}
\label{sec:curves}

A major breakthrough provided by Swift was the discovery of complex light curves at X-ray energies that led to introduction of the so-called canonical light curve \cite{nousek06,zhang06}, characterized by four decay segments: 1) an ``early steep decay'' with power-law decay indices $\sim$3 or even steeper; 2) a ``shallow or flat decay'' with a typical index around 0.5; 3) a normal decay, with indices around $1.2$; 4) finally, the late steep decay has typical values around 2. In $\sim50$\% of cases, flares are also  superimposed \cite{chincarini07,falcone07}. Some complexity is attributed to on-going central engine activity producing different emission components that originate from spatially distinct regions but that are temporally coincident, but alternative explanations invoking an external origin have also been suggested \cite{dermer07,dermer08}. Unlike other high-energy phenomena such as Active Galactic Nuclei, GRBs will remain spatially unresolved with current and future instrumentation. Model-dependent temporal properties therefore provide a powerful, indirect probe of the expanding fireball and its interaction with the surrounding medium. Polarisation measurements provide a {\em direct} probe.

At optical wavelengths, a variety of light curve properties are expected depending on the relative contributions of emission from reverse and forward shocks, the presence of additional energy injection by a long-lived central engine and the time of the observations with respect to the initial burst. 
\begin{figure}
\centering
\includegraphics[height=6cm]{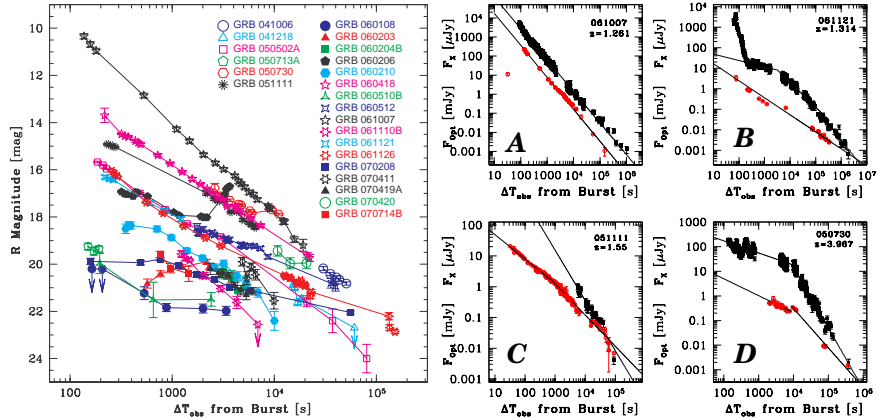}
%
%
\caption{{\em Left:} selection of early-time optical light curves from the Liverpool and Faulkes telescopes. {\em Right:} examples of X-ray (black) and optical (red) light curve comparisons leading to classification scheme using synchrotron model breaks \cite{melandri08}. }
\label{fig:melcurves}       
\end{figure}

Melandri et al.\cite{melandri08} classified a sample of 63 bursts via their optical and X-ray light curves. 50\% of the sample remained optically 'dark' despite rapid, deep observations through red filters. Optical counterparts ranged in brightness from R$\sim$10 mag to $\sim$22~mag in the first minutes after the burst and showed a range of decay behaviours (Fig. \ref{fig:melcurves}). Flares, although less common in the optical than X-ray band, were present and in some cases (e.g. GRB~060206) showed direct evidence of  significant energy injection \cite{monfardini06,guidorzi07,greiner09b,kruehler09}.
 
\begin{figure}
\centering
\includegraphics[height=4.1cm]{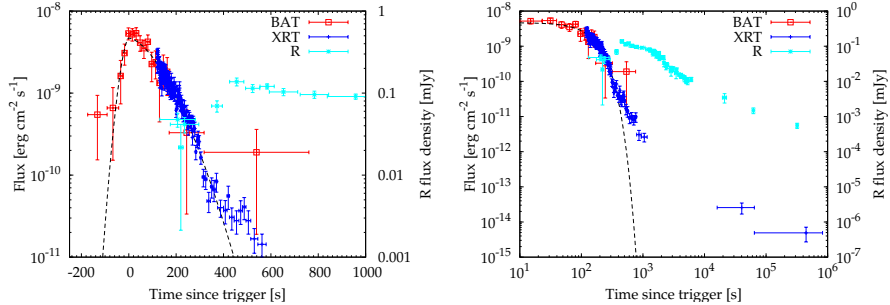}
%
%
\caption{$\gamma$-ray, X-ray and optical light curves of GRB~070419A. Left: early time light curves plotted on linear-log scale. Right: later time evolution included and shown on log-log scale (see Melandri et al.~\cite{melandri09a}).}
\label{fig:GRB070419A}       
\end{figure}

By comparing optical and X-ray light curves observed from t=100~s to $\sim$10$^{6}$~s post-burst, Melandri et al.~\cite{melandri08} introduced a coherent classification of optical/X-ray light curves under the framework of the standard fireball model and synchrotron theory, using the presence or absence of temporal breaks in each band (Fig. \ref{fig:melcurves}); the temporal location and evolution of chromatic and achromatic breaks depend on typical synchrotron and cooling frequencies with respect to the observing bands. Simply, class A shows no break in optical or X-ray bands because $\nu_c$ lies above or between the two bands; class B shows a break in the X-ray but not in the optical band as $\nu_c$ passes through the X-ray band; class C shows a break in the optical but not the X-ray band due to $\nu_c$ passing through the optical band and class D has breaks in both bands as energy injection stops or a jet break occurs. 60\% of optically detected bursts were explained with the forward shock model, while the remainder required energy injection and/or an ambient density gradient. GRB~070419A required a long-lived central engine ($\sim$250~s in $\gamma$ and X-rays), a finely tuned energy injection rate and an abrupt cessation of injection (Fig. \ref{fig:GRB070419A}).

\subsection{Long-Lived Central Engines Explain Bright and `Dark' Bursts}

Large robotic telescopes such as the LT provide deep upper limits at early times, (e.g. R$<$22~mag), thereby ruling out slow followup as a reason for non-detection of optical afterglows. Observing with red filters, the LT and FTs can identify GRBs at z$<$4 within the first few minutes after the burst via R-band photometric dropouts; a deep, early upper limit in R-band plus an i'-band detection quickly identified GRB~060927 as a moderately high-redshift source (z=5.467) \cite{guidorzi06b,velasco07}. Whilst a small fraction of optically undetected GRBs may lie at very high redshift so have their rest frame emission redshifted out of optical observing bands (e.g. GRB 090423 at z=8.1 \cite{tanvir09},\cite{salvaterra09}), others require a physical explanation for their darkness. Some may be explained by a relatively flat optical-to-X-ray spectral index and modest dust extinction (e.g. GRB~060108 \cite{oates06}; for typical extinction values, see, e.g., \cite{kann06,kann07}).  The underlying physical mechanism may be energy injection; as discussed above, a long-lived central engine is required to explain the properties of bright bursts but such a model may also provide a self-consistent explanation for dark bursts at intermediate redshift (z$<$6). Enhanced X-ray emission in the early time X-ray afterglow due to late-time central engine activity would make the bursts brighter than expected in the X-ray band compared to the optical band \cite{melandri08}. 

\section{Magnetized Fireballs?}
\label{sec:magnets}

The production of synchrotron radiation requires the presence of a magnetic field but the degree of magnetization and the configuration of the magnetic fields remain a matter of debate (e.g, \cite{lyutikov03,fan04,fan08}). Determination of the magnetic field properties is key to understanding the driving mechanism of the explosion, namely whether the ultrarelativistic ejecta are dominated by kinetic (baryonic) or magnetic (Poynting flux) energy. The ratio of magnetic and kinetic energy flux, $\sigma$, is used to express the magnetization (e.g. \cite{zhang05}) and is large if the magnetic field originates at the central engine and is advected outwards with the relativistic flow (Poynting flux jet). For a baryonic jet,  $\sigma$$\ll$ 1 and magnetic fields are assumed to be produced locally in the shock.  Magnetization may be inferred from optical light curves, given a variety of assumptions, or determined directly from measurements of the degree of polarisation of the optical afterglow. In both cases, observations at the earliest possible time after the burst - when the magnetic properties of the expanding fireball are still encoded in the emission - are essential to address these issues.

\paragraph{Indirect Observational Tests - Light Curve Evolution}

\begin{figure}
\centering
\includegraphics[height=6cm]{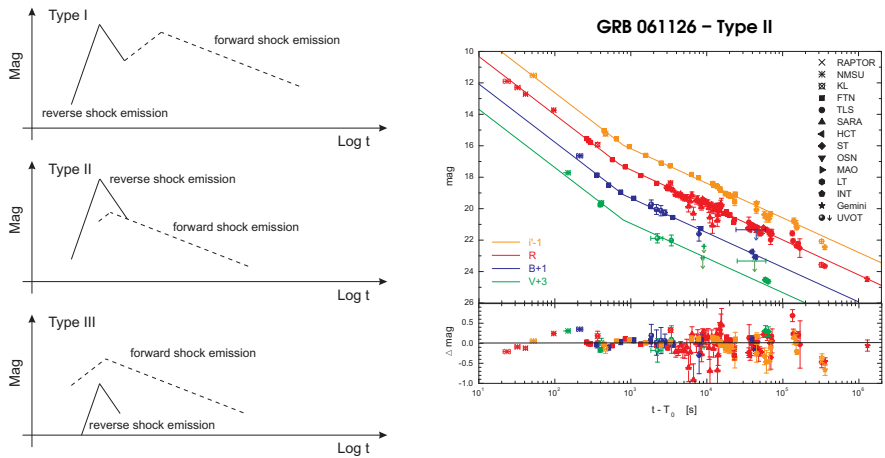}
%
%
\caption{{\em Left:}  Light-curves with different reverse-forward shock contributions. Magnetic energy density determines the relative contributions \cite{gomboc09}. {\em Right:} optical light curves of GRB~061126 showing the steep-to-shallow temporal break characteristic of a Type II reverse-forward shock observed after the reverse-shock peak\cite{gomboc08}. }
\label{fig:revshock}       
\end{figure}

In the standard fireball model (see \cite{piran99,zhang04,meszaros06,zhang07,piran07} for reviews) a shell of the relativistically expanding fireball collides with the circumburst medium to produce an external shock that results in the long-lived afterglow emission detectable from X-ray to optical, IR or sometimes radio frequencies. A reverse shock propagating backwards through the shell may, in  some circumstances, produce short-lived but bright optical emission - the so-called optical flash.
Figure~\ref{fig:revshock} shows three possible optical light curves with the relative contributions from the reverse and forward shocks identified. The strength of the reverse shock depends on magnetization and whether the typical frequency is close to the optical band \cite{fan04,zhang05}.  

\begin{figure}
\includegraphics[width=12cm]{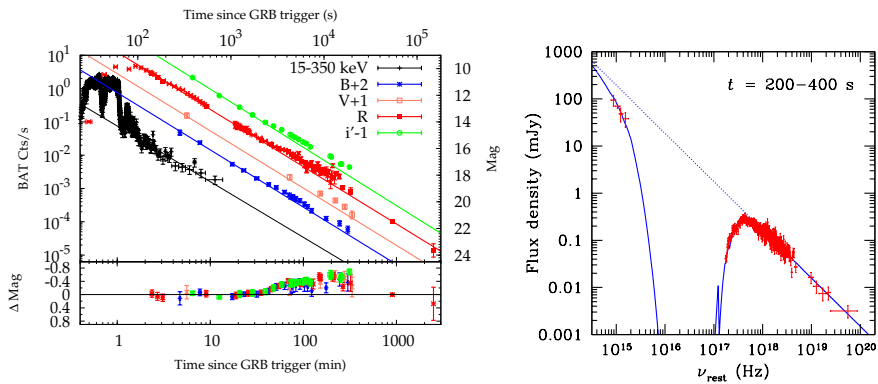} 
\caption{{\em Left:} $\gamma$-ray (black) and optical BVRi$'$ (colour) light curves of GRB~061007 all decaying with a steep power-law index $\alpha$$\sim$1.7 from 137~s to $\sim$10$^6$~s. Despite its optical brightness (R$\sim$10.3~mag at 137s), a reverse-shock optical flash is ruled out (adapted from \cite{mundell07_apj,rykoff09}). {\em Right:} early time (200-400~s) broad-band optical to $\gamma$-ray SED fitted with single absorbed power law $\beta$(opt-X-$\gamma$)=1.02$\pm$0.05 and rest-frame extinction $A_V(SMC)=0.48\pm0.19$~mag (from \cite{mundell07_apj}).}
\label{fig:GRB061007}       
\end{figure}

The dearth of reverse-shock optical flashes from Swift GRBs may be explained in the standard model via weak, non-relativistic reverse shocks and a typical frequency far below the optical band at early time; the bright optical counterpart to GRB~061007 (Fig.  \ref{fig:GRB061007}) illustrates this with its light curve comprising forward shock emission only and the typical frequency of the reverse shock emission being shown to lie in the radio band as early as 137~s after the burst \cite{mundell07_apj}. Alternatively, some optical flashes may be suppressed by strong magnetization. 

\paragraph{Direct Observational  Tests - Polarisation}

Synchrotron radiation is intrinsically highly polarised; specific properties of the emitting region may reduce the observed degree of polarisation (e.g. \cite{sari99,gruzinov99,ghisellini99,waxman03,greiner03,granot03,rossi04,sagiv04,gotz09}) so detection of significant polarisation at early time provides a direct signature of large-scale ordered magnetic fields in the expanding fireball \cite{lazzati04}.  Claims of high levels of polarisation observed in the gamma-ray prompt emission have remained controversial: Coburn \& Boggs \cite{coburn03} reported a high degree P=$80\pm20)$\% of  linear polarisation of the prompt emission of GRB~021206, but reanalysis of the data showed null polarisation \cite{rutledge04,wigger04}. In other cases, BATSE data of two bursts were reported to show evidence for P$>35$\% and $>50$\% using a GEANT4 model of the Earth albedo \cite{willis05}. More recently, observations of GRB~041219A with the INTEGRAL imager IBIS show polarization up to 43\% with rapid variability of degree and position angle \cite{gotz09},  although independent confirmation with spectrometer SPI data
remains difficult due to instrumental systematics \cite{mcglynn07,kalemci07}. In the remaining cases, only upper limits to possible large polarisation degrees have been reported (e.g. \cite{mcglynn09}).
Although careful calibration of instrumental systematics for $\gamma$-ray data remains challenging, these measurements show tantilising support for large-scale ordered magnetic fields in the region responsible for the prompt $\gamma$-ray production. Measuring the early optical polarisation provides a direct probe of the magnetic field configuration. Optical polarisation observations of GRBs have the advantage that stars in the GRB field of view can be used as simultaneous independent checks on any instrumental systematics. Theoretical models of GRBs predict that mildly magnetized outflows produce strong reverse shock emission that should be polarised \cite{lyutikov03,granot03,rossi04,fan04,fan05,covino07}, making early time optical polarisation measurements vital for direct determination of the magnetic field structure.

\begin{figure}
\centering
\includegraphics[height=5.2cm]{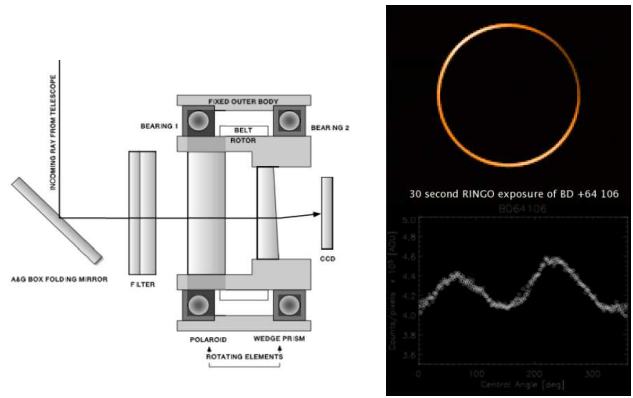}
\caption{{\em Left panel:} Diagram illustrating design of RINGO.
{\em Right panels:} Polarised star BD~+64~106 observed through RINGO (top panel);
calibrated trace around ring showing polarisation signature (from \cite{steele06}).}
\label{fig:RINGO}       
\end{figure}

The time-variable nature of optical emission from GRBs, however, makes traditional polarisation instruments unsuitable; we therefore designed the novel RINGO polarimeter on the LT \cite{steele06}, whose overall layout is shown in Fig.~\ref{fig:RINGO}.
Its design was based on a novel ring polarimeter concept explored by Clarke \& Neumayer \cite{clarke02}.
This makes use of a rotating Polaroid to modulate the incoming polarised flux to be studied and is followed by deviating optics
which co-rotate. A filter equivalent to V$+$R bands was chosen to optimise the signal to noise ratio,
as estimated from the typical GRB colours and to avoid the fringing on the CCD that the I filter would have caused.
The result of this design is that each source image is recorded on the CCD as a ring, with the polarisation signal mapped out
along the ring in a $\sin{(2\,\theta)}$ pattern (right panels of Fig.~\ref{fig:RINGO}). One benefit of this design
is the reduction of the saturation constraints on high precision photometry, as the flux is spread the flux over the pixels
along the ring. The potential problem of many rings overlapping with each other is only a concern for crowded fields which correlate with low galactic latitudes, i.e. where GRB optical counterparts are strongly extinguished and therefore observationally already disfavoured. The design was optimised through requirements on i) the rotation speed, to minimise impact on the polarisation ring profiles; ii) the geometry of the deflecting optics, to optimise the ring size. The polarimetric accuracy of RINGO is
a few percents for a $R=15$-mag object in a 10-s long exposure and decreases to a few 0.1\% in a 10-minutes exposure.
Further details on the RINGO polarimeter and its calibration and data reduction can be found in Steele et al. \cite{steele06}.

\begin{figure}
\centering
\includegraphics[height=5.2cm]{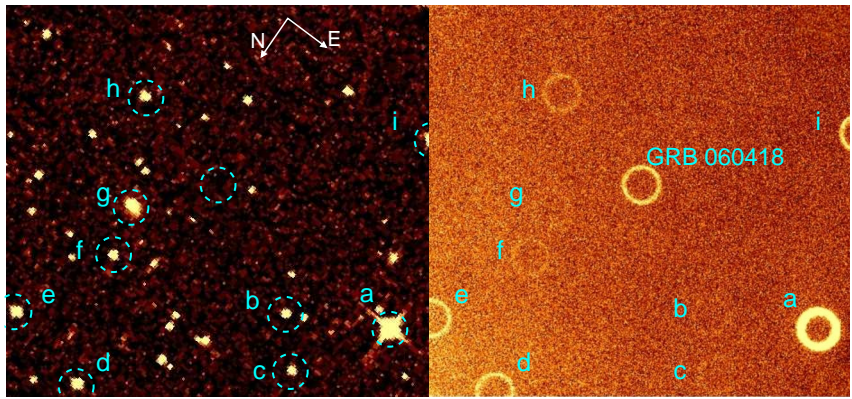}
\caption{{\em Left:} DSS image of the field before the occurrence of GRB~060418.
{\em Right:} RINGO image showing GRB~060418 and other field sources as rings (from \cite{mundell07_sci}.)}
\label{fig:RINGO_060418}       
\end{figure}

\begin{figure}
\centering
\includegraphics[width=12cm]{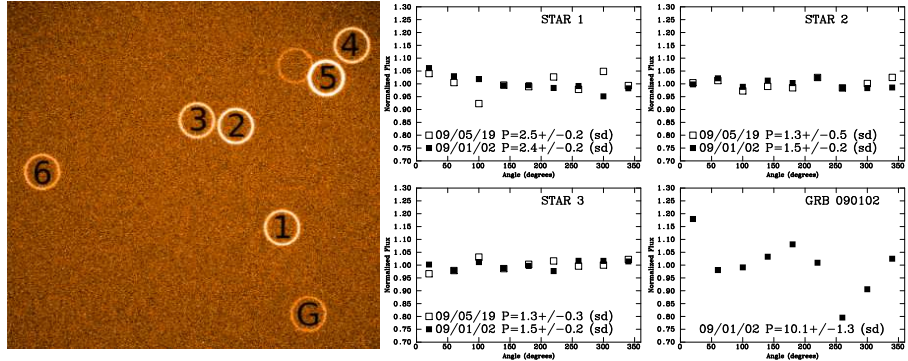} 
\caption{{\em Left:} RINGO image of field containing GRB~090102 (marked 'G'); field stars are also visible. {\em Right:} Traces around rings for stars 1, 2, 3 and GRB 090102; derived polarisation \% for each is quoted. \cite{steele09}.}
\label{fig:GRB090102}       
\end{figure}

\begin{figure}
\centering
\includegraphics[height=5.0cm]{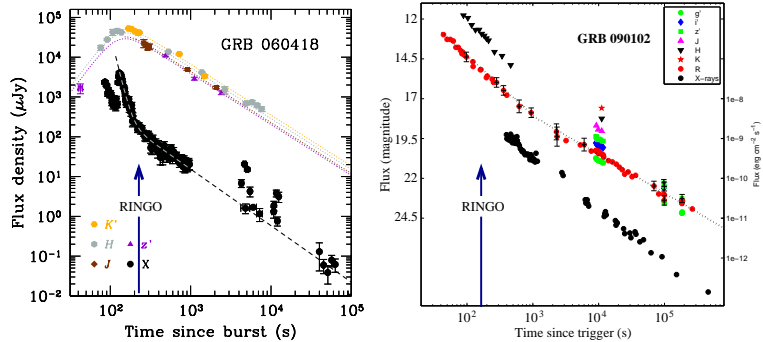}
\caption{Light curves of GRB~060418 and GRB~090102 with time of RINGO observation marked at 203~s and 160~s (adapted from \cite{molinari07} and \cite{gendre09}) respectively.}
\label{fig:lcs0418_0102}       
\end{figure}

RINGO has now measured the optical polarisation of two GRBs (Figs.~\ref{fig:RINGO_060418}, \ref{fig:GRB090102}, \ref{fig:lcs0418_0102}): Mundell et al. (2007)  reported  {\bf {\em P$<$8\%}}  in GRB~060418 at 203~s post-burst \cite{mundell07_sci} and Steele et al. 2009 measured {\bf {\em P~=~10.2$\pm$1.3\%}} in GRB~090102 at 160.8~s post-burst \cite{steele09}. In the standard GRB fireball model both a forward shock, propagating into the circumburst medium, and a reverse shock, propagating into the GRB ejecta,  contribute to the observed afterglow emission \cite{kobayashi99,gomboc09} (Fig.~\ref{fig:revshock}). The measurement in GRB~060418 was made close to the peak in the optical light curve at the time of deceleration of the fireball associated with the onset of the afterglow, when the reverse and forward shocks contributed equally to the detected light, while GRB~090102 was measured when the light curve was dominated by reverse-shock emission.  Below we summarise the observations and suggest a unifying model to reconcile the two results.

\paragraph{Optical polarisation measurements of GRB~060418 and GRB~090102}

The RINGO images of the fields containing GRB~060418 and 090102 are shown in Figs.~\ref{fig:RINGO_060418}
 and \ref{fig:GRB090102}. GRB~060418 was detected by the Swift Burst Alert Telescope (BAT) on April 18, 2006 and showed a multi-pulsed gamma-ray profile with a total duration of 52~s, followed by a separate small bump at 130~s concomitant with a large X-ray flare observed with the Swift X-Ray Telescope (XRT) \cite{falcone06a,falcone06b} and probably due to prolonged internal activity (see Fig. \ref{fig:lcs0418_0102}) .
A number of robotic facilities reacted to the prompt Swift alert and detected the optical and NIR counterpart: in particular, REM detected a smooth optical rise peaking at 153~s, interpreted as the onset of the afterglow \cite{molinari07}, whereas the X-ray curve was seen to decline from the beginning, apart from the large flare mentioned above (Fig.~\ref{fig:lcs0418_0102}). The LT was triggered automatically, starting observations with a 30-s long exposure taken with the RINGO polarimeter at 203~s post-GRB i.e. simultaneously with the fading tail of the prompt gamma-ray emission
(Fig.~\ref{fig:lcs0418_0102}) and the close to the X-ray flare. However, the contribution in the optical band from the X-ray flare was estimated to be negligible, thus confirming that RINGO measured the afterglow \cite{mundell07_sci}.  

Swift detected GRB 090102 on 2 January 2009 at 02:55:45 UT, with a pulse of gamma rays lasting T90 =27 s and comprising four overlapping peaks starting 14-s before the trigger time \cite{sakamoto09}. The automatic localization was communicated to ground-based facilities, and a single 60-second RINGO exposure was obtained starting 160.8 seconds after the burst. Simultaneous with our polarization observation of GRB 090102, a number of automated photometric follow-ups were performed by other facilities \cite{steele09}.  The optical light curve (Fig.~\ref{fig:lcs0418_0102}) beginning at 40-s postburst, is fitted by a steep-shallow broken power law $\alpha_1$ = 1.50$\pm$0.06 then $\alpha_2$ = 0.97$\pm$0.03 after approximately 1000~s \cite{gendre09}. In contrast, the X-ray light curve began at 396 s post burst and shows a steady decay consistent with a single power law with slope $\alpha$ = 1.36$\pm$0.01 and no evidence of flares or breaks up to t$>$7$\times$10$^5$~s \cite{gendre09}. 
The absence of additional emission components, e.g. from late-time central engine activity, in the optical light curves of GRB 060418 and 090102 allows a straightforward interpretation of their early time polarisation in the context of current GRB models. 

Despite the brightness of the optical afterglow of GRB~060418, no dominant reverse shock was observed, similar to other cases in which the typical frequency is inferred to lie below the optical band at early time such as GRB~061007 \cite{mundell07_apj}. Instead, the light curve peak is typical of the forward shock. The forward shock peaks either when fireball decelerates or when the typical synchrotron frequency crosses the observed band: the latter was ruled out for GRB~060418, as no colour evolution was observed around the peak, thus supporting the deceleration interpretation with the synchrotron frequency lying below the NIR bands at the peak time \cite{molinari07}. This is also consistent with the steep ($|\alpha|\sim2.7$) rise observed in the NIR bands (Fig.~\ref{fig:lcs0418_0102}), in agreement with the theoretical expectations for the forward shock deceleration \cite{kobayashi03}. The RINGO measurement of GRB~060418 was made close to the deceleration time at the onset of the afterglow and when any magnetic field would still be present in the emitting region because the detected light contains an equal contribution from forward and reverse shocks, i.e. Type III \cite{melandri08, gomboc09} (Fig.~\ref{fig:revshock}). 

\begin{figure}
\centering
\includegraphics[width=11cm]{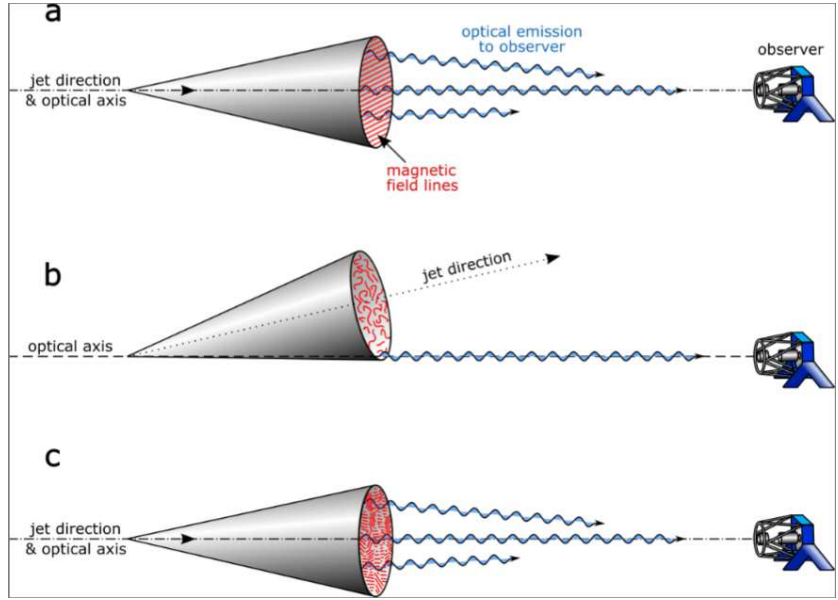} 
\caption{Competing models of GRB magnetic field structure for standard fireball model: (a) large-scale ordered magnetic field  (favoured for GRB~090102); (b) tangled magnetic field in shock front and line of sight close to jet edge, and (c) shock front contains some independent patches of locally-ordered magnetic fields \cite{steele09}.}
\label{fig:GRB090102_schem}       
\end{figure}

In contrast to GRB~060418, GRB~090102 exhibited the steep-shallow optical decay typical of that expected from an afterglow whose emission is dominated by the reverse shock emission at early times (Fig.~\ref{fig:lcs0418_0102}); this is the first GRB for which polarized optical light at early time has been detected and its high level of polarisation {\em P=10.2$\pm$1.3\%}  requires the presence of large-scale ordered magnetic fields in the outflow \cite{steele09} (Fig.~\ref{fig:GRB090102_schem}).  As the RINGO measurement was made when the reverse shock emission dominated the light curve, the large polarisation signal provides the first {\em direct} evidence that reverse shocks are produced in the presence of such fields.

Figure~\ref{fig:GRB090102_schem} shows three competing models of GRB magnetic field structure for the standard fireball model that  could produce polarised light: (a) large-scale ordered magnetic field threaded through the GRB outflow - this model is favoured for GRB~090102; (b) tangled magnetic field generated in the shock front and an off-axis line-of-sight close to the jet edge - a scenario ruled out by the optical light curves for which an off-axis jet would require a shallow-to-steep decay, where the opposite is observed in GRB~090102; (c) a shock front containing independent patches of locally-ordered magnetic fields - a scenario that is unlikely to account for GRB~090102 because the {\em maximum} predicted polarisation is ~10\% under the extreme assumption that the coherence length grows at the speed of light in the local fluid frame after the field is generated.

In summary, the polarisation properties of GRB~060418  and GRB~090102 can be reconciled in a single model in which the outflow is mildly magnetised ($\sigma$$\sim$1) and contains large-scale ordered magnetic fields. In GRB~060418, $\sigma$ slightly larger than unity is needed to suppress the reverse shock emission (and hence polarisation), while in GRB~090102, $\sigma$ slightly less than unity will produce the bright polarised reverse shock emission that is observed. 

\section{Summary and Prospects}
Exploring the extreme physics exhibited by GRB explosions is technically challenging due to (a) the unpredictability of their occurrence (b) their short-lifetime rapidly fading emission (c) the wide range of observed brightnesses of optical counterparts, ranging from R~=~5 to $>$22~mag within minutes of the burst itself - all drivers of autonomous follow-up technology. Deep, fast and multi-filter observations are crucial to identify the counterparts to these events that represent the brightest stellar objects observed out to the epoch of reionization.  With the advent of Swift,  discoveries such as the canonical early X-ray decay, the X-ray flares, the detection of the afterglows of short-duration GRBs,  and the recognition that GRB-producing stars exist out at least to z=8.2 keep GRB studies at the forefront of astrophysics.

Efforts continue to understand the complexity of the X-ray versus optical afterglow temporal evolution, the circumburst environment properties, in particular the dust versus gas content around the GRB progenitor and along the line of sight through the host galaxy, and the origin(s) of optical flares and their possible interpretations (e.g. GRB~080129 \cite{greiner09b}: e.g  residual collisions in the GRB outflow or hot spots in strongly magnetized ejecta?\cite{lyutikov06}). Questions remain on the fundamental nature of the relativistic ejecta, the underlying radiation mechanisms and the role of magnetic fields. Observational surprises such as the relative lack of GeV emission from many bursts detected by Fermi and the rich variety in optical properties of GRB counterparts continue to drive developments in GRB modelling and observational technology. The 2-m robotic telescopes described here are proving decisive in tackling many of these issues: the variety of light curves has been investigated both on statistical grounds \cite{melandri08} as well as in individual cases of special interests: the dark GRB population as well as the luminosity function distribution have been characterised over a broad range of apparent brightnesses and the presence and lack of reverse shock emission in specific cases was investigated.  

The interplay between forward and reverse shocks within the standard fireball model, as determined by the magnetic properties of the outflow, may yet succeed in explaining the dearth of reverse shocks previously expected from pre-Swift observations.  The use of the RINGO polarimeter on the LT, capable of measuring the polarisation of optical counterparts to GRBs as early as a few minutes after the onset of the prompt $\gamma$-ray emission, has provided the earliest measurements and detection of GRB polarisation, setting important direct constraints on the magnetic field structure of the fireball and on the jet configuration \cite{mundell07_sci,steele09}.  Further progress made in understanding the magnetic field structure of the fireball - large-scale ordered fields are currently preferred over locally tangled fields in the shock layer- will contribute to our knowledge of the nature of the outflow along the jets.  Time-resolved early time polarisation light curves (\% and PA) from the newly commissioned RINGO2 promise to provide unprecedented diagnostic information on the structure and evolution of the outflow and its magnetic field for a statistically significant sample of GRBs down to R$<$17~mag and thus allow powerful discrimination between predictions of the hydrodynamical versus magnetised jet models and ultimately constrain the physics of GRB central engines. 

{\it Acknowledgment.}
We thank the Liverpool Telescope group for technical, scientific and artwork support.
CGM acknowledges financial support from Research Councils U.K. The Liverpool Telescope is located at the Observatorio
del Roque de Los Muchachos, La Palma, Canary Islands, Spain. It was designed and built by Telescope Technologies
Ltd and is owned and operated by Liverpool John Moores University with financial support from the U.K. PPARC.
The Faulkes telescopes, owned by Las Cumbres Observatory, are located in Hawaii (FTN) and Siding Spring, Australia (FTS)
and are operated with support from the Dill Faulkes Educational Trust.


\begin{thebibliography}{}
\bibitem{klebesadel73} R.~W. Klebesadel, I.~B.~Strong, R.~A. Olson, ``Observations
  of gamma-ray bursts of cosmic origin'', ApJ, 182, pp. L85-L88, 1973.
\bibitem{costa97} E.~Costa, F.~Frontera, J. Heise, et al., ``Discovery of an X-ray
afterglow associated with the $\gamma$-ray burst of 28 February 1997'', Nature,
vol. 387, pp. 783-785, 1997.
\bibitem{vanparadijs97} J. Van Paradijs, P.~J. Groot, T. Galama, et al.,
``Transient optical emission from the error box of the $\gamma$-ray burst of 28 February 1997'',
Nature, vol. 386, pp. 686-689, 1997.
\bibitem{frail97} D.~A. Frail, S.~R. Kulkarni, L.~Nicastro, et al., ``The radio afterglow
from the $\gamma$-ray burst of 8 May 1997'', Nature, vol. 389, pp. 261-263, 1997.
\bibitem{gehrels04} N. Gehrels, G.~Chincarini, P.~Giommi, et al., ``The Swift gamma-ray
burst mission'', ApJ, vol. 611, pp. 1005-1020, 2004.
\bibitem{paciesas99} W.~S. Paciesas, C.~A. Meegan, G.~N. Pendleton, et~al.,
``The fourth BATSE gamma-ray burst catalog (revised)'', ApJS, vol. 122, pp. 465-495, 1999.
\bibitem{barthelmy08} S.~D.~Barthelmy, ``GCN and VOEvent: A status report'', Astron. Nachr.,
vol. 329, pp. 340-342, 2008.
\bibitem{akerlof99} C.~Akerlof, R.~Balsano, S.~Barthelmy, et al.,
``Observation of contemporaneous optical radiation from a gamma-ray burst'', Nature,
vol. 398, pp. 400-402, 1999.
\bibitem{boella97} G.~Boella, R.~C.~Butler, G.~C.~Perola, et al., ``BeppoSAX, the wide band mission
for X-ray astronomy'', A\&A Suppl. Ser., vol. 122, pp. 299-307, 1997.
\bibitem{frail01} D.~A. Frail, S.~R. Kulkarni, R.~Sari, et al., 
``Beaming in Gamma-Ray Bursts: Evidence for a Standard Energy Reservoir'', ApJ, vol. 562,
pp. L55-L58, 2001.
\bibitem{ghirlanda04} G.~Ghirlanda, G.~Ghisellini, D. Lazzati, C. Firmani,
  ``Gamma-ray bursts: new rulers to measure the universe'', ApJ, vol. 613, pp. L13-L16, 2004.
\bibitem{melandri08} A. Melandri, C.~G. Mundell, S. Kobayashi, et al.,
``The early-time optical properties of gamma-ray burst afterglows'', ApJ, vol. 686,
pp. 1209-1230, 2008.
\bibitem{cenko09} S.~B. Cenko, J. Kelemen, F.~A. Harrison, et al.,
``Dark bursts in the Swift era: the palomar 60 inch - Swift early optical afterglow catalog'',
ApJ, vol. 693, pp. 1484-1493, 2009.
\bibitem{oates09} S.~R. Oates, M.~J. Page, P.~Schady, et~al., ``A statistical study of
gamma-ray burst afterglows measured with the Swift Ultraviolet Optical Telescope'', MNRAS,
vol. 395, pp. 490-503, 2009.
\bibitem{yost07} S.~A.~Yost, F.~Aharonian, C.~W.~Akerlof, et al., ``The dark side of ROTSE-III
prompt GRB observations'', ApJ, vol. 669, pp. 1107-1114, 2007. 
\bibitem{roming06} P.~W.~A. Roming, P. Schady, D.~B. Fox, et al., ``Very Early Optical
Afterglows of Gamma-Ray Bursts: Evidence for Relative Paucity of Detection'', ApJ,
vol. 652, pp. 1416-1422, 2006.
\bibitem{tagliaferri05} G.~Tagliaferri, M.~Goad, G.~Chincarini, et al.,
``An unexpectedly rapid decline in the X-ray afterglow emission of long $\gamma$-ray bursts'',
Nature, vol. 436, pp. 985-988, 2005.
\bibitem{burrows05} D.~N. Burrows, P.~Romano, A.~Falcone, et al., ``Bright X-ray flares in gamma-ray
burst afterglows'', Science, vol. 309, pp. 1833-1835, 2005.
\bibitem{obrien06} P.T.~O'Brien, R.~Willingale, J.~Osborne, et al.,
``The early X-ray emission from GRBs'', ApJ, vol. 647, pp. 1213-1237, 2006.
\bibitem{chincarini07} G.~Chincarini, A.~Moretti, P.~Romano, et al.,
``The First Survey of X-Ray Flares from Gamma-Ray Bursts Observed by Swift: Temporal Properties
and Morphology'', ApJ, vol. 671, pp. 1903-1920, 2007.
\bibitem{falcone07} A. D. Falcone, D. Morris, J. Racusin, et al., ``The First Survey of X-ray Flares
from Gamma-Ray Bursts Observed with Swift: Spectral Properties and Energetics'', ApJ, vol. 671,
pp. 1921-1938, 2007.
\bibitem{klotz09} A.~Klotz, M.~Bo\"er, J.~L.~Atteia, B.~Gendre, ``Early optical observations
of gamma-ray bursts by the TAROT telescopes: period 2001--2008'', AJ, vol. 137, pp. 4100-4108, 2009.
\bibitem{rykoff09} E.~S.~Rykoff, F.~Aharonian, C.~W.~Akerlof, et al.,
``Looking into the fireball: ROTSE-III and Swift observations of early GRB afterglows'',
arXiv:0904.0261, submitted.
\bibitem{graham09} J. F. Graham, A.~S. Fruchter, A.~J. Levan, et al.,
``GRB 070714B: discovery of the highest spectroscopically confirmed short burst redshift'',
ApJ, vol. 698, pp. 1620-1629, 2009.
\bibitem{kawai06} N.~Kawai, G.~Kosugi, K.~Aoki, ``An optical spectrum of the afterglow
of a $\gamma$-ray burst at a redshift of z=6.295'', Nature, vol. 440, pp. 184-186, 2006.
\bibitem{greiner09a} J.~Greiner, T.~Kr\"uhler, J.~P.~U. Fynbo, et al., ``GRB~080913 at
redshift 6.7'', ApJ, vol. 693, pp. 1610-1620, 2008.
\bibitem{salvaterra09} R.~Salvaterra, M.~Della Valle, S.~Campana, et al., ``GRB~090423 at a
redshift of z$\approx$8.1'', Nature, vol. 461, pp. 1258-1260, 2009.
\bibitem{tanvir09} N.~R.~Tanvir, D.~B.~Fox, A.~J.~Levan, ``A $\gamma$--ray burst at a redshift
of z$\approx$8.2'', Nature, vol. 461, pp. 1254-1257, 2009.
\bibitem{zerbi01} R.~M.~Zerbi, G.~Chincarini, G.~Ghisellini, et al., ``The REM telescope: detecting
the near infra-red counterparts of Gamma-Ray Bursts and the prompt behavior of their optical continuum'',
Astron. Nachr., vol. 322, pp. 275-285, 2001.
\bibitem{vestrand04} W.~T.~Vestrand, K. Borozdin, D.~J.~Casperson, et al., ``RAPTOR: Closed-Loop monitoring
of the night sky and the earliest optical detection of GRB 021211'',Astron. Nachr., vol. 325, pp. 549-552, 2004. 
\bibitem{reichart05} D.~Reichart, M.~Nysewander, J. Moran, et al., ``PROMPT: Panchromatic Robotic Optical
Monitoring and Polarimetry Telescopes'', Il Nuovo Cimento C, vol. 28, pp. 767-770, 2005. 
\bibitem{yost06} S.~A.~Yost, F.~Aharonian, C.~W.~Akerlof, et al., ``Status of the ROTSE-III telescope
network'', Astron. Nachr., vol. 327, pp. 803-805, 2006.
\bibitem{klotz08} A.~Klotz, M.~Bo\"er, J. Eysseric, et al.,
``Robotic Observations of the Sky with TAROT: 2004-2007'', PASP, vol. 120, pp. 1298-1306, 2008.
\bibitem{cenko06} S.~B.~Cenko, D.~B.~Fox, D.-S. Moon, et al.,
  ``The automated Palomar 60 Ich Telescope'', PASP, vol. 118, pp. 1396-1406, 2006.
\bibitem{bloom06} J.~S. Bloom, D.~L. Starr, C.H. Blake et al., ``Autonomous Observing and Control
Systems for PAIRITEL, a 1.3m Infrared Imaging Telescope'', ASP Conf. Ser., vol. 351, pp. 751-754, 2006.
\bibitem{guidorzi06a} C.~Guidorzi, A.~Monfardini, A.~Gomboc, et al.,
``The automatic real-time gamma-ray burst pipeline of the 2-m Liverpool Telescope'',
PASP, vol. 118, pp. 288-296, 2006.
\bibitem{nousek06} J.~Nousek, C.~Kouveliotou, D.~Grupe, et al.,
``Evidence for a Canonical Gamma-Ray Burst Afterglow Light Curve in the Swift XRT Data'',
ApJ, vol. 642, pp. 389-400, 2006.
\bibitem{zhang06} B.~Zhang, Y.~Z.~Fan, J.~Dyks, et al., ``Physical processes shaping gamma-ray burst
X-ray afterglow light curves: theoretical implications from the Swift X-ray Telescope observations'',
ApJ, vol. 642, pp. 354-370, 2006.
\bibitem{dermer07} C. D. Dermer, ``Rapid X-ray declines and plateaus in Swift GRB light curves explained
by a highly radiative blast wave'', ApJ, vol. 664, pp. 384-396, 2007.
\bibitem{dermer08} C. D. Dermer, ``Nonthermal synchrotron radiation from gamma-ray burst external
shocks and the X-ray flares observed with Swift'', ApJ, vol. 684, pp. 430-448, 2008.
\bibitem{monfardini06} A.~Monfardini, S.~Kobayashi, C.~Guidorzi, et al., ``High-quality early-time
light curves of GRB~060206: implications for gamma-ray burst environments and energetics'',
ApJ, vol. 648, pp. 1125-1131, 2006.
\bibitem{guidorzi07} C.~Guidorzi, S.~D.~Vergani, S.~Sazonov, et al., ``GRB~070311: a direct link
between the prompt emission and the afterglow'', A\&A, vol. 474, pp. 793-805, 2007.
\bibitem{greiner09b} J.~Greiner, T.~Kr\"uhler, S.~McBreen, et al., ``A strong optical flare
before the rising afterglow of GRB~080129'', ApJ, vol. 693, pp. 1912-1919, 2009.
\bibitem{kruehler09} T.~Kr\"uhler, J.~Greiner, S.~McBreen, et al.,
  ``Correlated optical and X-ray flares in the afterglow of XRF~071031'', ApJ, vol. 697, pp. 758-768, 2009.
\bibitem{melandri09a} A.~Melandri, C.~Guidorzi, S.~Kobayashi, et al., ``Evidence for energy injection
and a fine-tuned central engine at optical wavelengths in GRB~070419A'', MNRAS, vol. 395, 1941-1494, 2009.
\bibitem{guidorzi06b} C.~Guidorzi, D.~Bersier, A.~Melandri, et al., ``GRB~060927: Faulkes Telescope South observation'', GCN Circulars, 5633, http://gcn.gsfc.nasa.gov/gcn3/5633.gcn3
\bibitem{velasco07} A.~E.~Ruiz-Velasco, H.~Swan, E.~Troja, et al., ``Detection of GRB~060927 at z=5.47:
implications for the use of gamma-ray bursts as probes of the end of the dark ages'', ApJ, vol. 669,
pp. 1-9, 2007.
\bibitem{oates06} S.~R.~Oates, C.~G.~Mundell, S.~Piranomonte, et al., ``Anatomy of a dark burst:
the afterglow of GRB~060108'', MNRAS, vol. 372, pp. 327-337, 2006.
\bibitem{kann06} A.~Kann, S.~Klose, A.~Zeh, ``Signatures of extragalactic dust in pre-Swift GRB afterglows'',
ApJ, vol. 641, pp. 993-1009, 2006.
\bibitem{kann07} A.~Kann, S.~Klose, B.~Zhang, et al., ``The afterglows of Swift-era gamma-ray bursts. I.
Comparing pre-Swift and Swift long/soft (Type II) GRB optical afterglows'', arXiv:0712.2186, submitted.
\bibitem{lyutikov03} M.~Lyutikov, V.~I.~Pariev, R.~D.~Blandford, ``Polarization of Prompt Gamma-Ray
Burst Emission: Evidence for Electromagnetically Dominated Outflow'', ApJ, vol. 597, pp. 998-1009, 2003.
\bibitem{fan04} Y.-Z.~Fan, D.-M.~Wei, C.~F.~Wang, ``The very early afterglow powered by
ultra-relativistic mildly magnetized outflows'', A\&A, vol. 424, pp. 477-484, 2004.
\bibitem{fan08} Y.-Z.~Fan, D.~Xu, D.-M.~Wei, ``Polarization evolution accompanying the very
early sharp decline of gamma-ray burst X-ray afterglows'', MNRAS, vol. 387, pp. 92-96, 2008.
\bibitem{zhang05} B.~Zhang, S.~Kobayashi, ``Gamma-ray burst early afterglows: reverse shock emission
from an arbitrarily magnetized ejecta'', ApJ, vol. 628, pp. 315-334, 2005.
\bibitem{piran99} T.~Piran, ``Gamma-ray bursts and the fireball model'', Phys. Rep., vol. 314,
pp. 575-667, 1999.
\bibitem{zhang04} B.~Zhang, P.~Meszaros, ``Gamma-ray bursts: progress problems and prospects'',
Int. J. Mod. Phys. A, vol. 19, pp. 2385-2472, 2004.
\bibitem{meszaros06} P.~Meszaros, ``Gamma-ray bursts'', Rep. Prog. Phys., vol. 69, pp. 2259-2321, 2006.
\bibitem{zhang07} B.~Zhang, ``Gamma-Ray Bursts in the Swift Era'', Chin. J. Astron. Astrophys.,
vol. 7, pp. 1-50, 2007.
\bibitem{piran07} T.~Piran, ``Gamma-ray burst theory after Swift'', Phil. Trans. R. Soc. A, vol. 365,
pp. 1151-1162, 2007.
\bibitem{gomboc09} A.~Gomboc, S.~Kobayashi, C.~G.~Mundell, et al., ``Optical flashes, reverse shocks
and magnetization'', AIP Conference Proceedings,  vol. 1133, pp. 145-150, 2009.
\bibitem{gomboc08} A.~Gomboc, S.~Kobayashi, C.~Guidorzi, et~al., ``Multiwavelength analysis of
the intriguing GRB~061126: the reverse shock scenario and magnetization'', ApJ, vol. 687,
pp. 443-455, 2008.
\bibitem{mundell07_apj} C.~G. Mundell, A.~Melandri, C.~Guidorzi, et al., ``The remarkable afterglow
of GRB~061007: implications for optical flashes and GRB fireballs'', ApJ, vol. 660, pp. 489-495, 2007.
\bibitem{sari99} R.~Sari, ``Linear polarization and proper motion in the afterglow of beamed
gamma-ray bursts'', ApJ, vol. 524, pp. L43-L46, 1999.
\bibitem{gruzinov99} A. Gruzinov, E. Waxman, ``Gamma-ray burst afterglow: polarization and analytic
light curves'', ApJ, vol. 511, pp. 852-861, 1999.
\bibitem{ghisellini99} G.~Ghisellini, D.~Lazzati, ``Polarization light curves and position angle variation
of beamed gamma-ray bursts'', MNRAS, vol. 309, pp. L7-L11, 1999.
\bibitem{waxman03} E.~Waxman, ``New directions for $\gamma$-rays'', Nature, vol. 423, pp. 388-389,
2003.
\bibitem{greiner03} J.~Greiner, S.~Klose, K.~Reinsch, ``Evolution of the polarization of the optical afterglow of the $\gamma$-ray burst GRB030329'', Nature, vol. 426, pp. 157-159, 2003.
\bibitem{gotz09} D.~G\"otz, P. Laurent, F. Lebrun, et al., ``Variable polarization measured
in the prompt emission of GRB~041219A using IBIS on board INTEGRAL'', ApJ, vol. 695,
pp. L208-L212, 2009.
\bibitem{lazzati04} D.~Lazzati, S.~Covino, J.~Gorosabel, et al., ``On the jet structure and magnetic
field configuration of GRB~020813'', A\&A, vol. 422, pp. 121-128, 2004.
\bibitem{granot03} J.~Granot, A.~K\"onigl, ``Linear Polarization in Gamma-Ray Bursts: The Case
for an Ordered Magnetic Field'', ApJ, vol. 594, pp. L83-L87, 2003
\bibitem{rossi04} E.~M.~Rossi, D.~Lazzati, J.~D.~Salmonson, ``The polarization of afterglow
emission reveals $\gamma$-ray bursts jet structure'', MNRAS, vol. 354, pp. 86-100, 2004.
\bibitem{fan05} Y.~Z.~Fan, B.~Zhang, D.~M.~Wei, ``Early Optical-Infrared Emission from GRB 041219a:
Neutron-rich Internal Shocks and a Mildly Magnetized External Reverse Shock'', ApJ, vol. 628,
pp. L25-L28, 2005.
\bibitem{covino07} S.~Covino, ``A Closer Look at a Gamma-Ray Burst'', Science, vol. 315,
pp. 1798-1798, 2007.
\bibitem{coburn03} W.~Coburn, S.~E.~Boggs, ``Polarization of the prompt $\gamma$-ray emission
from the $\gamma$-ray burst of 6 December 2002'', Nature, vol. 423, pp. 415-417, 2003.
\bibitem{rutledge04} R.~E.~Rutledge, D.~B.~Fox, ``Re-analysis of polarization in the $\gamma$-ray
flux of GRB~021206'', MNRAS, vol. 350, pp. 1288-1300, 2004.
\bibitem{wigger04} C.~Wigger, W.~Hajdas, K.~Arner et al., ``Gamma-Ray Burst Polarization: Limits from
RHESSI Measurements'', ApJ, vol. 613, pp. 1088-1100, 2004. 
\bibitem{willis05} D.~R.~Willis, E.~J.~Barlow, A.~J.~Bird, et al., ``Evidence of polarisation in the
prompt gamma-ray emission from GRB~930131 and GRB~960924'', A\&A, vol. 439, pp. 245-253, 2005.
\bibitem{kalemci07} E.~Kalemci, S.~E.~Boggs, C.~Kouveliotou, et al., ``Search for polarization
from the prompt gamma-ray emission of GRB~041219A with SPI on INTEGRAL'', ApJS, vol. 169,
pp. 75-82, 2007.
\bibitem{mcglynn07} S.~McGlynn, D.~J.~Clark, A.~J.~Dean, et al., ``Polarisation studies of the
prompt gamma-ray emission from GRB~041219A using the spectrometer aboard INTEGRAL'', A\&A,
vol. 466, pp. 895-904, 2007.
\bibitem{mcglynn09} S.~McGlynn, S.~Foley, B.~McBreen, et al., ``High energy emission and
polarisation limits for the INTEGRAL burst GRB~061122'', A\&A, vol. 499, pp. 465-472, 2009.
\bibitem{steele06} I.~A.~Steele, S.~D.~Bates, D.~Carter, et al., ``RINGO: a novel ring polarimeter
for rapid GRB followup'', Proceedings of the SPIE, vol. 6269, pp. 62695M, 2006.
\bibitem{clarke02} D.~Clarke, D.~Neumayer, ``Experiments with a novel CCD stellar explosions'',
A\&A, vol. 383, pp. 360-366, 2002. 
\bibitem{falcone06a} A.~D.~Falcone, S.~D.~Barthelmy, D.~N.~Burrows, et al., ``GRB~060418: 
Swift detection of a burst with bright x-ray and optical afterglow'', GCN Circulars, 4966,
http://gcn.gsfc.nasa.gov/gcn3/4966.gcn3
\bibitem{falcone06b} A.~D.~Falcone, D.~N.~Burrows, J.~Kennea, ``GRB~060418: Swift XRT Team
Refined Analysis'', GCN Circulars, 4973, http://gcn.gsfc.nasa.gov/gcn3/4973.gcn3
\bibitem{molinari07} E.~Molinari, S.~D.~Vergani, D.~Malesani, et al.,
``REM observations of GRB~060418 and GRB~060607A: the onset of the afterglow and the initial
fireball Lorentz factor determination'', A\&A, vol. 469, pp. L13-L16, 2007.
\bibitem{mundell07_sci} C.~G. Mundell, I.~A.~Steele, R.~J.~Smith, et al., ``Early optical polarization
of a gamma-ray burst afterglow'', Science, vol. 315, pp. 1822-1824, 2007.
\bibitem{sakamoto09} T.~Sakamoto, S.~D.~Barthelmy, W.~H.~Baumgartner et al., ``GRB~090102: Swift-BAT
refined analysis'', GCN Circulars, 8769, http://gcn.gsfc.nasa.gov/gcn3/8769.gcn3.
\bibitem{gendre09} B.~Gendre, A.~Klotz, E.~Palazzi et al., ``Testing GRB models with the strange afterglow of GRB~090102'', arXiv:0909.1167, 2009.
\bibitem{kobayashi99} S.~Kobayashi, T.~Piran, R. Sari, ``Hydrodynamics of a relativistic fireball:
the complete evolution'', ApJ, vol. 513, pp. 669-678, 1999.
\bibitem{kobayashi03} S.~Kobayashi, B.~Zhang, ``GRB~021004: reverse shock emission'', ApJ, vol. 582,
pp. L75-L78, 2003.
\bibitem{sagiv04} A.~Sagiv, E.~Waxman, A.~Loeb, ``Probing the Magnetic Field Structure in Gamma-Ray
Bursts through Dispersive Plasma Effects on the Afterglow Polarization'', ApJ, vol. 615, pp. 366-377, 2004.
\bibitem{lyutikov06} M.~Lyutikov, ``The electromagnetic model of gamma-ray bursts'', New J. Phys.,
vol. 8, pp. 119, 2006.
\bibitem{steele09} I.~A.~Steele, C.~G.~Mundell, R.~J.~Smith et al., ``Ten per cent polarized optical emission from GRB~090102'', Nature, vol. 462, pp. 767-769, 2009.
\end{thebibliography}
\end{document}